\documentclass[%
	final,
  conferencejournal,
    comsoc,
	letterpaper,
	oneside,
	twocolumn,
	nofonttune,%
]{IEEEtran}%
\PassOptionsToPackage{usenames,dvipsnames,svgnames,x11names,table,prologue}{xcolor}%
\PassOptionsToPackage{hyphens}{url}%
\PassOptionsToPackage{nomessages}{fp}%
\ifCLASSINFOpdf
  \usepackage[pdftex]{graphicx}
\else
  \usepackage[dvips]{graphicx}
\fi
\ifCLASSOPTIONcompsoc
   \usepackage[caption=false,font=normalsize,labelfont=sf,textfont=sf]{subfig}
\else
   \usepackage[caption=false,font=footnotesize]{subfig}
\fi

\usepackage{stfloats}
\usepackage[english]{babel}%
\usepackage[utf8]{inputenc}%
\usepackage[babel,style=english]{csquotes}%
\usepackage{hyphsubst}%
\usepackage[%
	activate={true,%
	nocompatibility},%
	final,%
	tracking=true,%
	kerning=true,%
	spacing=true,%
	factor=1100,%
	stretch=10,%
	shrink=10%
]{microtype}%
\usepackage{setspace}%
\usepackage{xcolor}%
\definecolor{todonotecol}{RGB}{250,0,0}%
\usepackage{xparse}
\usepackage[%
	colorlinks=false,%
	urlcolor=black,%
	linkcolor=black,%
	citecolor=black,%
	filecolor=black,%
	breaklinks,%
	]{hyperref}%
\usepackage{url}%
\usepackage[%
	acronym,%
	nopostdot,%
	seeautonumberlist,%
	shortcuts,%
	section=chapter,%
	toc,%
]{glossaries}%
\glsaddkey*{url}
{}
{\glsentryurl}
{\Glsentryurl}
{\glsurl}
{\Glsurl}
{\GLSurl}
\glsaddkey*{longpltwo}
{}
{\glsentrylongpltwo}
{\Glsentrylongpltwo}
{\glslongpltwo}
{\Glslongpltwo}
{\GLSlongpltwo}
\newglossaryentry{}{%
	name={},%
	plural={},%
	description={},%
}%
\newacronym{dmz}{DMZ}{demilitarized zone}
\newacronym{it}{IT}{information technology}
\newacronym{ot}{OT}{operational technology}
\newacronym{opcua}{OPC UA}{Open Platform Communications Unified Architecture}
\newacronym{tsn}{TSN}{time-sensitive networking}
\newacronym{scada}{SCADA}{supervisory control and data acquisition}
\newacronym{vpc}{VPC}{Virtual Process Controller}
\newacronym{plc}{PLC}{programmable logic controller}
\newacronym{udp}{UDP}{User Datagram Protocol}
\newacronym{pubsub}{PubSub}{Publish/Subscribe}
\newacronym{vm}{VM}{virtual machine}
\newacronym{pc}{PC}{personal computer}
\newacronym{ptp}{PTP}{Precision Time Protocol}
\newacronym{ntp}{NTP}{Network Time Protocol}
\newacronym{splc}{Soft-PLC}{Software-PLC}
\newacronym{rte}{RTE}{Real-Time Ethernet}
\newacronym{osi}{OSI}{Open Systems Interconnection}
\newacronym{mac}{MAC}{Media Access Control}
\newacronym{e2e}{E2E}{end-to-end}
\newacronym{tcp}{TCP}{Transmission Control Protocol}
\newacronym{hmi}{HMI}{Human-Machine Interface}
\newacronym{kpi}{KPI}{Key Performance Indicator}
\newacronym{mes}{MES}{Manufacturing Execution System}
\newacronym{rami4.0}{RAMI 4.0}{Reference Architectural Model Industrie 4.0}
\newacronym{tacnet4.0}{TACNET 4.0}{Tactile Internet 4.0}
\newacronym{3gpp}{3GPP}{3rd Generation Partnership Project}
\newacronym{5gppp}{5GPPP}{5G Infrastructure Public Private Partnership}
\newacronym{itu}{ITU}{International Telecommunication Union}
\newacronym{ngnm}{NGNM}{Next Generation Mobile Networks}
\newacronym{agv}{AGV}{automated guided vehicle}
\newacronym{v2v}{V2V}{vehicle-to-vehicle}
\newacronym{v2x}{V2X}{vehicle-to-infrastructure}
\newacronym{d2x}{D2X}{device-to-x}
\newacronym{qos}{QoS}{quality of service}
\newacronym{noa}{NOA}{NAMUR Open Architecture}
\newacronym{dcs}{DCS}{distributed control system}
\newacronym{m2m}{M2M}{machine-to-machine}
\newacronym{sdn}{SDN}{software-defined networking}
\newacronym{erp}{ERP}{enterprise resource planning}
\newacronym{ip}{IP}{Internet Protocol}
\newacronym{lte}{LTE}{Long Term Evolution}
\newacronym{5g}{5G}{5th generation wireless communication system}
\newacronym{harq}{HARQ}{Hybrid Automatic Repeat Request}
\newacronym{rat}{RAT}{radio access technology}
\newacronym{mc}{MC}{Multi-Connectivity}
\newacronym{fbmc}{FBMC}{Filter Bank Multicarrier}
\newacronym{gfdm}{GFDM}{Generalized Frequency Division Multiplexing}
\newacronym{sla}{SLA}{service-level agreement}
\newacronym{5qi}{5QI}{5G QoS Indicator}
\newacronym{bmbf}{BMBF}{German Federal Ministry of Education and Research}
\newacronym{nrt}{NRT}{non real-time}
\newacronym{irt}{IRT}{isochronous real-time}
\newacronym{rt}{RT}{real-time}
\newacronym{ar}{AR}{augmented reality}
\newacronym{wlan}{WLAN}{wireless local area network}
\newacronym{cpu}{CPU}{central processing unit}
\newacronym{ram}{RAM}{random-access memory}
\newacronym{ue}{UE}{user equipment}
\newacronym{ran}{RAN}{radio access network}
\newacronym{bs}{BS}{base station}
\newacronym{cn}{CN}{core network}
\newacronym{epc}{EPC}{evolved packet core}
\newacronym{mec}{MEC}{mobile edge cloud}
\newacronym{rtt}{RTT}{round-trip time}
\newacronym{vpn}{VPN}{virtual private network}
\newacronym{vpf}{VPF}{Virtual Process Control Function}
\newacronym{vdi}{VDI}{Association of German Engineers}
\newacronym{ipc}{IPC}{industrial pc}
\newacronym{rtos}{RTOS}{real-time operating system}
\newacronym{vpcmo}{VPC M\&O}{VPC Management \& Orchestration}
\newacronym{i/o}{I/O}{input /output}
\newacronym{fbd}{FBD}{Function Block Diagram}
\newacronym{ldd}{LDD}{Ladder Logic Diagram}
\newacronym{sfc}{SFC}{Sequential Function Chart}
\newacronym{il}{IL}{Instruction List}
\newacronym{st}{ST}{Structured Text}
\newacronym{os}{OS}{operating system}
\newacronym{ir}{IR}{Inactive Resource}
\newacronym{msc}{MSC}{message sequence chart}
\newacronym{icps}{ICPS}{industrial cyber-physical system}
\newacronym{cps}{CPS}{cyber-physical systems}
\newacronym{mtd}{MTD}{Moving Target Defense}
\newacronym{iot}{IoT}{Internet of Things}
\newacronym{iiot}{IIoT}{industrial Internet of Things}
\newacronym{iiot2}{IIoT}{Industrial IoT}
\newacronym{k8s}{K8s}{Kubernetes}
\newacronym{rkt}{rkt}{CoreOS Rocket}
\newacronym{dcos}{DC/OS}{Distributed Cloud Operating System}
\newacronym{ml}{ML}{machine learning}
\newacronym{ai}{AI}{artificial intelligence}
\newacronym{ict}{ICT}{industrial communication technology}
\newacronym{kvm}{KVM}{kernel-based virtual machine}
\newacronym{ie}{IE}{Industrial Ethernet}
\newacronym{ias}{IAS}{industrial automation system}
\newacronym{cots}{COTS}{commercial off-the-shelf}
\newacronym{slat}{SLAT}{Second Level Address Translation}
\newacronym{gui}{GUI}{graphical user interface}
\newacronym{l2}{L2}{layer 2}
\newacronym{l3}{L3}{layer 3}
\newacronym{ouc}{OUC}{Open User Communication}
\newacronym{api}{API}{application programming interface }
\newacronym{fb}{FB}{function block}
\newacronym{mqtt}{MQTT}{message queuing telemetry transport }
\newacronym{amqp}{AMQP}{advanced message queuing protocol}
\newacronym{embb}{eMBB}{enhanced mobile broadband}
\newacronym{urllc}{URLLC}{ultra-reliable low-latency communication}
\newacronym{mmtc}{mMTC}{massive machine-type communications}
\newacronym{npn}{NPN}{non-public network}
\newacronym{prb}{PRB}{Physical Ressource Block}
\newacronym{gnb}{gNB}{5G base station}
\newacronym{5gs}{5GS}{5G system}
\newacronym{5gc}{5GC}{5G core network}
\newacronym{oai}{OAI}{OpenAirInterface}
\newacronym{enb}{eNB}{4G base station}
\newacronym{sdr}{SDR}{Software Defined Radio}
\newacronym{usrp}{USRP}{Universal Software Radio Peripheral}
\newacronym{mme}{MME}{Mobility Management Entity}
\newacronym{hss}{HSS}{Home Subscriber Server}
\newacronym{pdn}{PDN}{Packet Data Network}
\newacronym{sgw}{SGW}{Serving Gateway}
\newacronym{gtp}{GTP}{GPRS Tunneling Protocol}
\newacronym{mimo}{MIMO}{Multiple Input Multiple Output}
\newacronym{fpga}{FPGA}{Field Programmable Gate Array}
\newacronym{pss}{PSS}{primary synchronization signal}
\newacronym{sss}{SSS}{secondary synchronization signal}
\newacronym{sfn}{SFN}{system frame number}
\newacronym{mib}{MIB}{master information block}
\newacronym{pbch}{PBCH}{Physical Broadcast Channel}
\newacronym{ism}{ISM}{Industrial, Scientific and Medical}
\newacronym{e-utra}{E-UTRA}{evolved UMTS Terrestrial Radio Access}
\newacronym{bldc}{BLDC}{brushless DC}
\newacronym{dl}{DL}{Downlink}
\newacronym{ds-tt}{DS-TT}{Device-Side TSN Translator}
\newacronym{nw-tt}{NW-TT}{Network-Side TSN Translator}
\newacronym{upf}{UPF}{User Plane Function}
\newacronym{tsi}{TSi}{ingress timestamping}
\newacronym{tse}{TSe}{egress timestamping}
\newacronym{gm}{GM}{grandmaster}
\newacronym{soap}{SOAP}{Simple Object Access Protocol}
\newacronym{http}{HTTP}{Hypertext Transfer Protocol}
\newacronym{pdu}{PDU}{protocol data unit}
\newacronym{coap}{CoAP}{Constrained Application Protocol}
\newacronym{lqr}{LQR}{linear–quadratic regulator}
\newacronym{pid}{PID}{proportional–integral–derivative}
\glsdisablehyper
\usepackage[%
	backend=biber,%
	style=ieee,%
	isbn=false,%
	hyperref=true,%
	maxbibnames=99,%
	sorting=none,%
	natbib=true,%
	language=english,%
	defernumbers=true,%
	]{biblatex}%
\DeclareFieldFormat{sentencecase}{\csname bbx@colon@search\endcsname#1}

\usepackage{nameref}
\usepackage{soul}
\usepackage{balance}
\usepackage{textcomp}
\usepackage{calc}
\usepackage{xkeyval}
\usepackage{multirow}
\usepackage{tabularx}
\usepackage{tabulary}
\usepackage{makecell}
\usepackage{verbatim}%
\usepackage{pgfplots}
\usepackage{tikz}
\usepackage{textcomp} 
\newcommand{\nl}{\par\noindent} 
\newcommand{\mytilde}{{\raise.17ex\hbox{$\scriptstyle\mathtt{\sim}$}}}

\newlength\textheighttemp%
\newlength\textwidthtemp%
\newlength\textheightstd%
\newlength\textwidthstd%
\newlength\textheightold%
\newlength\textwidthold%
\newlength\tempheight%
\newlength\tempwidth%
\SetProtrusion{encoding={*},family={bch},series={*},size={6,7}}
              {1={ ,750},2={ ,500},3={ ,500},4={ ,500},5={ ,500},
               6={ ,500},7={ ,600},8={ ,500},9={ ,500},0={ ,500}}
\SetExtraKerning[unit=space]
    {encoding={*}, family={bch}, series={*}, size={footnotesize,small,normalsize}}
    {\textendash={400,400}, 
     "28={ ,150}, 
     "29={150, }, 
     \textquotedblleft={ ,150}, 
     \textquotedblright={150, }} 
\SetTracking{encoding={*}, shape=sc}{40}
\makeatletter
\let\blx@rerun@biber\relax
\makeatother

\usepackage{textcomp}
\usepackage{newtxmath}
\usetikzlibrary{external}
\pgfplotsset{
  grid style = {
   line width = 0.1pt
  }
}
\definecolor{blue}{RGB}{0,68,170}%
				\newcommand{\disablewr}[1]{#1}%
				\newcommand{\newcommanddisw}[3]{\newcommand{#1}[1]{\disablewr{\textcolor{#2}{#3}}}}%
\renewcommand{\disablewr}[1]{}%
\newcommand{\hide}[1]{}

\definecolor{todocol}{named}{red}
\newcommanddisw{\todo}{todocol}{ToDo: #1}%
\definecolor{migucol}{named}{purple}%
\newcommanddisw{\migucom}{migucol}{{@}comment: #1}%
\newcommanddisw{\miguhigh}{migucol}{#1}%
\definecolor{josccol}{named}{brown}%
\newcommanddisw{\josccom}{josccol}{{@}comment: #1}%
\newcommanddisw{\joschigh}{josccol}{#1}%

\begin{document}%
%


\IEEEoverridecommandlockouts
\title{Assessing Open Interfaces and Protocols of PLCs for Computation Offloading at Field Level%
\thanks{This research was supported by the German Federal Ministry for Economic Affairs and Energy (BMWi) within the project FabOS under grant number 01MK20010C. The responsibility for this publication lies with the authors. This is a preprint of a work accepted but not yet published at the 17th IEEE International Conference on Factory Communication Systems (WFCS). Please cite as: M. Gundall and H.D. Schotten: “Assessing Open Interfaces and Protocols of PLCs for Computation Offloading at Field Level”. In: 2021 17th IEEE International Conference on Factory Communication Systems (WFCS), IEEE, 2021.}
       }

%
%
\author{%
\IEEEauthorblockN{%
    Michael Gundall\IEEEauthorrefmark{1} %
    and Hans D. Schotten\IEEEauthorrefmark{1}\IEEEauthorrefmark{2} %
    \\%
}%
\IEEEauthorblockA{%
    \IEEEauthorrefmark{1}German Research Center for Artificial Intelligence GmbH (DFKI), Kaiserslautern, Germany \\%
    \IEEEauthorrefmark{2}Department of Electrical and Computer Engineering, Technische Universität Kaiserslautern, Kaiserslautern, Germany %
	\\%
    Email: %
        \{michael.gundall, hans\_dieter.schotten%
        \}@dfki.de
        \\%
}%
}%


%

%
%
%
%
%
%
%
%
\maketitle
%
%
%
%
%
\begin{abstract}%
\Glspl{plc} are the core element of industrial plants in todays deployments. They read sensor values, execute control algorithms, and write output values. 
Furthermore, industrial plants have lifetimes of one or more decades. Thus, in a realistic Industry 4.0 scenario, these devices have to be integrated in novel systems. In order to apply advanced concepts and technologies, such as computation offloading, which requires data exchange between \glspl{plc} and edge cloud, 
 we investigate open communication interfaces of two typical \glspl{plc} of Siemens S7 series. 
Hence, each of the interfaces is analyzed based on plug \& play capability, if metadata is provided, protocol efficiency, and performance. For the latter, the smallest possible update time for each of the interfaces will be measured. 
\end{abstract}%
%
%
%
%
%
\IEEEpeerreviewmaketitle
%
%
%
%
%
%
%
%
\section{Introduction}%
\label{sec:Introduction}

The convergence of \gls{it} and \gls{ot} is one important enabler for realizing Industry 4.0 use cases \cite{gundall20185g}, whereby \gls{5g} is seen as key technology for realizing mobile use cases \cite{gundall2021introduction}. In addition, mobile devices, such as \glspl{agv} or drones, can profit by the so-called computation offloading, e.g., to save energy  \cite{8254628}. Thus, if these devices provide a high mobility, e.g. movement between factory halls, the offloaded algorithms also have to be mobile  \cite{gundall2020introduction,gundall2021feasability}. 
However, there are reasons for not only using computation offloading for mobile devices. Since algorithms are getting more and more complex, the computational power of resource constrained devices, such as \glspl{plc} may be exceeded. Furthermore, Industry 4.0 describes “lot size one" what requires a reconfiguration of process controllers at very short time intervals. Since legacy devices do not provide this flexibility but are required as interface to sensors and actuators due to life-cycle-times of industrial plants of ten years or more \cite{5535166}, offloading of algorithms to edge clouds is a suitable approach. 

In order to realize data exchange between \glspl{plc} and devices that are located in the area of the \gls{ot} and applying novel technologies, such as virtualization \cite{gundall2020application}, open interfaces and protocols are required. Therefore, we investigate communication protocols of \glspl{plc} that are natively compatible with \gls{ot} hardware and thus using standard Ethernet and \gls{ip} layer.

\section{Accessing Data from PLCs}
\label{sec:comp offloading}

In this section the possibilities for accessing data of two \glspl{plc} are examined, which are part of the S7 series. 
Therefore, the individual interfaces are described in detail and a comparison is carried out (see Tab. \ref{tab1}). Besides the specification of the protocol, qualitative aspects, such as plug \& play capability and the availability of metadata is investigated. In addition, the protocol efficiency, which can be expressed by the payload divided by the total number of bytes sent, is determined for 1, 10, and 100 data values, where each data value is assumed to be 4 bytes. Furthermore, the update time plays a major role as it indicates the frequency with which data packets can be sent. It is defined as the \textit{“[...] time interval between any two consecutive messages delivered to the application."} \cite{3gppts22104}. Therefore, we examine the minimum update time that the device can use to send a new data packet for 1, 10, and 100 data values. This value is characteristic for the investigated \gls{plc} and network independent. 


\begin{table*}[tb]
\caption{Comparison of the interfaces of the investigated \glspl{plc} using standard Ethernet.}
\begin{center}
\begin{tabular*}{\textwidth}{|c|c|c|c|c|p{0.04\textwidth}|p{0.04\textwidth}|p{0.04\textwidth}|p{0.04\textwidth}|p{0.04\textwidth}|p{0.04\textwidth}|p{0.04\textwidth}|p{0.04\textwidth}|p{0.04\textwidth}|}
\cline{1-14}
\multicolumn{2}{|c|}{\textbf{Interface}}  & \textbf{Protocol} & \textbf{Plug \&} & \textbf{Meta-}  & \multicolumn{9}{|c|}{\textbf{Protocol Efficiency \& Min. Update Time}} \\
\cline{6-14}
\multicolumn{2}{|c|}{\textbf{Configuration}} &   & \textbf{Play} & \textbf{data} &   \multicolumn{3}{|c|}{\textbf{1}} & \multicolumn{3}{|c|}{\textbf{10}} & \multicolumn{3}{|c|}{\textbf{100}}\\
\multicolumn{2}{|c|}{} & &  &   & \multicolumn{3}{|c|}{\textbf{Data Value}} & \multicolumn{3}{|c|}{\textbf{Data Values}} & \multicolumn{3}{|c|}{\textbf{Data Values}}\\
\multicolumn{2}{|c|}{} & &    &  &  &\textbf{\textit{314}} &\textbf{\textit{1512}} & & \textbf{\textit{314}} &\textbf{\textit{1512}} & &\textbf{\textit{314}} &\textbf{\textit{1512}} \\
\multicolumn{2}{|c|}{} & &    &  & \textbf{[\%]} & \multicolumn{2}{|c|}{\textbf{[ms]}} & \textbf{[\%]} &\multicolumn{2}{|c|}{\textbf{[ms]}} &\textbf{[\%]} &\multicolumn{2}{|c|}{\textbf{[ms]}} \\
\cline{1-14} 
\multicolumn{2}{|c|}{Open User} & UDP & - & - &  5.6& 1.00 & 3.61 & 42.6 & 1.00 & 3.60 &88.1 & 1.00 & 3.63 \\
\cline{3-14} 
\multicolumn{2}{|c|}{Communication} & TCP & - & - &  2.8& 1.01 &3.77 & 22.5 & 1.04 & 3.78 & 74.4 &1.02  &3.83 \\
\cline{1-14} 
\multicolumn{2}{|c|}{LIBNODAVE} & ISO on TCP & + & - & 1.2 & 2.00 &1.32 & 10.8 &2.00 & 1.32 & 54.6$^{\mathrm{2}}$ &4.00 & 1.40\\
\cline{1-14} 
OPC & Write & UATCP & - & + &  0.9 & n/a&6.83 & 6.3 &n/a& 7.36 & 14.5 &n/a &16.56\\
 UA & Service &  &  &  &   & & & &&  & & &\\
\cline{2-14} 
Server & Read  & UATCP & + & + &  0.8 &n/a& 9.11 & 3.7 &n/a& 30.35 & 5.4 &n/a& 246.1\\
Client & Service &  &  &  &   & & & &&  & & &\\
\cline{1-14}
\multicolumn{2}{|c|}{OPC UA PubSub} & UADP & - & $\circ$ &  5.5 & n/a&1.02 & 33.9 & n/a & 1.26 & 70.4$^{\mathrm{1}}$ & n/a& 2.30$^{\mathrm{1}}$\\
\cline{1-14} 
\multicolumn{14}{l}{$^{\mathrm{1}}$As the current firmware of the PLC allows a maximum of 20 data values to be transferred, but it can be assumed that this} \\
\multicolumn{14}{l}{~~restriction will be removed in a newer version, the value was estimated.} \\
\multicolumn{14}{l}{$^{\mathrm{2}}$For the S7-314 series this value is halved, because two requests have to be sent.} \\
\end{tabular*}
\label{tab1}
\end{center}
\end{table*}

\begin{table}[htbp]
\caption{Overview about name and size of the different data messages.}
\begin{center}
\begin{tabular*}{\columnwidth}{|c|c|c|c|c|}
\cline{1-5}
\multicolumn{2}{|c|}{\textbf{Message Specification}} & \multicolumn{3}{|c|}{\textbf{Message Size [byte]}} \\
\multicolumn{2}{|c|}{\textbf{}} & \multicolumn{3}{|c|}{\textbf{Data Values}}    \\
\textbf{Interface} & \textbf{Name} & \textbf{1}  & \textbf{10}  & \textbf{100} \\
\cline{1-5}
Open User & \textit{UDP\_Data} & 72 & 94 & 454 \\
Communication & \textit{TCP\_Data} & 144 & 178 & 538 \\
 \cline{1-5}
LIBNODAVE & \textit{Job} & 169 & 169 & 169 \\
 & \textit{Ack\_Data} & 167 & 203 & 563 \\
\cline{1-5}
OPC  & \textit{WriteRequest} & 234 & 396 & 2154 \\
UA & \textit{WriteResponse} & 202 & 238 & 598 \\
Server&\textit{ReadRequest} & 270 & 756 & 5778 \\
Cient  &\textit{ ReadResponse} & 212 & 338 & 1598 \\
\cline{1-5}
OPC UA & \textit{DataSetMessage} & 73 & 118 & 568 \\
PubSub &  &  &  &  \\
\cline{1-5}
\end{tabular*}
\label{tab: Message size}
\end{center}
\end{table}

\subsection{\gls{ouc}}
The \gls{ouc} was originally developed with the intention to allow  multiple \glspl{plc} to exchange data using the following \gls{ip}-based protocols:
\begin{itemize}
\item \gls{udp} (RFC 768),
\item \gls{tcp} (RFC 793), and
\item ISO-on-\gls{tcp} (RFC 1006).
\end{itemize}
Since it cannot exclusively used by \glspl{plc}, 
this interface is well suited for offloading data to edge devices. As ISO-on-\gls{tcp}, which is also referred to as “S7~Protocol", does not bring an advantage compared to standard \gls{tcp}, it is not discussed in the \gls{ouc} section.

\subsubsection{\gls{udp}}
For sending data from the \gls{plc} to a mini PC using \gls{udp}-based \gls{ouc}, two \glspl{fb} (TCON, TUSEND) must be configured in the \gls{plc}. Among other things, the \gls{ip} address and the port number must be specified there. For this reason the \gls{plc} must be stopped to be able to use the new software module. Therefore, the \gls{ouc} does not allow plug \& play mechanisms to avoid downtime of the \gls{plc}. Also, the data to be sent is packed directly into the payload of the \gls{udp} packet, without any additional information. This means that no metadata is provided and the receiver must know exactly how the data is structured, such as the byte order and data type. Due to this fact, the payload is very heterogeneous, but a very efficient data transmission can be realized, since only the \gls{udp}, \gls{ip} and Ethernet II headers of 8, 20, and 26 bytes have to be added. This results in a protocol efficiency of 5.6\% for 1 data value, 42.6\% for 10 data values, and 88.1\% for 100 data values. One of the drawbacks is that the \glspl{plc} is not able to send a packet in each cycle. This results in a update time of $\approx$ 3.6~ms for S7-1512 \gls{plc}. However, the udpate time for the S7-314 \gls{plc} is significantly lower, compared to S7-1512 \gls{plc}. Here, update times of 1~ms are possible. Furthermore, the update times of both \glspl{plc} are almost independent of the payload size.

\subsubsection{\gls{tcp}}

The use of \gls{tcp} in \gls{ouc} is comparable to \gls{udp}. Besides the configuration of a different \gls{fb} (TSEND), the only difference lies in the transport protocol, as the name implies. Here the characteristic is that each of the packets is acknowledged and thus no packet loss can occur, since lost packets are automatically retransmitted. This results in a higher reliability, but also a higher overhead, compared to \gls{udp}. Thus, in addition to the larger header of \gls{tcp}, which is 20 bytes, the 72 bytes per acknowledge reduce the protocol efficiency, which is almost half for 1 and 10 data values and about  $15 \%$ lower for 100 data values compared to \gls{udp}. In addition, the larger header and acknowledgment generation increases the minimum update time slightly.

\subsection{LIBNODAVE}

LIBNODAVE is a free and open source library for using the ISO-on-\gls{tcp} protocol communicating on TCP/IP port 102 for data exchange with Siemens S7 \glspl{plc} \cite{hergenhahn2011libnodave}. If the RJ-45 port where the cable is connected is not explicitly disabled, any device supporting the S7 protocol can communicate directly with the \gls{plc}. This enables plug \& play capability, but is also the reason why this protocol has already been used for cyber attacks such as Stuxnet \cite{langner2013kill}. This means that appropriate security measures must be taken if there is a connection to the Internet or if there is a possibility that malware can be placed on a device that communicates with the \gls{plc}, because on some series not only can data values be read and written, but the complete \gls{plc} can also be stopped. This is especially true for older models like the 300 and 400 series. In the 1500 series, the function has been severely restricted so that critical functions such as start and stop can no longer be executed by any device. However, read and write access is still possible. This makes it a suitable protocol for accessing data of the \glspl{plc} studied in this paper.

Since in this communication method the S7 protocol is built on top of \gls{tcp}, the protocol overhead is larger compared to \gls{ouc}. In addition, the data exchange must be triggered by the edge device. Since all network traffic required for the data exchange should be considered, the messages required to query the data and its acknowledgements must also be considered. To request data, the so-called \textit{Job} message is sent. As shown in Tab. \ref{tab: Message size}, this remains the same size for both 1, 10, and 100 data values. This is because the data is requested over its memory area and the length of the data chunk. The response of the \glspl{plc}, called \textit{Ack\_Data} message, contains the data and consequently becomes larger, with increasing the number of data values. Considering both messages results in low protocol efficiencies of 1.2\%, 10.8\% and 54.6\% without any meta information about the sent data.  In addition, the \gls{pdu} size of the S7-314 \gls{plc} is limited to 240 bytes. Therefore, only $\approx$ 50 data values can be read within one request. In order to obtain 100 data values, two requests have to be send. This results in twice the update time, protocol overhead, and thus, in a halved protocol efficiency. However, until a maximum of polling 50 data values, the update time of the S7-314 \gls{plc} is constantly 2~ms. Regarding the S7-1512 \gls{plc}, the aforementioned restriction is not given. Furthermore, it is able to send a new data packet every 1.2, 1.32, and 1.4~ms. 

\subsection{\gls{opcua}}
To close the gap between \gls{it} and \gls{ot}, \gls{opcua} \cite{IEC625411} was introduced. It aims at a secure, simple and platform-independent exchange of information between industrial applications \cite{leitner2006opc}. For this purpose, it provides both a self-describing information model and various communication protocols. Even though the information model in conjunction with the data exchange is a major milestone in industrial automation, we will focus on the communication protocols in the following. Since \gls{opcua} is not supported for the S7-300 series, no values can be measured for this interface for the S7-314 \gls{plc}. If computation offloading using \gls{opcua} server client or \gls{pubsub} is explicitly required, a gateway has to be connected next to the \gls{plc}, as proposed in \cite{6009198}. To offload data from the S7-314 \gls{plc} to the gateway, the investigated interfaces can also be used. Then the data can be offloaded to an edge device.

\subsubsection{OPC UA Server Client}

The \gls{opcua} server client pattern supports the binary \gls{tcp}-based communication protocol (UATCP) as well as a solution that is well suited for web services based on \gls{soap}/\gls{http}. Due to lower resource consumption and less overhead, which is important for embedded devices like \glspl{plc}, we focus on the UATCP protocol running on port 4840.

Two different services are possible for data exchange between \gls{plc} and the edge node, depending on the role of each device in the specific scenario, since the client and server roles are strictly defined. The client sends requests to the server that are answered with a response. Thus, if the \gls{plc} is the client and wants to send data to the mini PC, it must send so-called \textit{WriteRequests} containing the data values to be written to the server's address space. Then, this message is replied with the \textit{WriteResponse} to give the client a response with some information, such as a status code. Also, both messages contain a lot of metadata, such as the timestamp of the device, the unique identifier of the variable in the address space, and the data type, just to name some of the information. This makes the protocol very powerful, but comes with a larger overhead. Therefore, the \textit{WriteRequests} have the sizes of 234 bytes, 396 bytes, and 2154 bytes for 1, 10, and 100 data values, respectively. In addition, the \textit{WriteResponses} add another 202 bytes, 238 bytes, and 598 bytes. This results in a low protocol efficiency that is, for example, $\mathrm{<1\%}$ for 1 data value.

If the roles are swapped, so the \gls{plc} being the server and the remote application running on the edge node being the client, the data must be exchanged via the \textit{ReadService}. This means that the edge node is the client and must poll the data from the server.  To do this, a \textit{ReadRequest} containing the variables to be read is sent to the server running on the \gls{plc}. The message is then responded with the \textit{ReadResponse}, which contains the current data values. Similar to the \textit{WriteService}, the data also contains a lot of metadata. Therefore, the protocol efficiency is also low for 1, 10, and 100 data values. Due to the protocol overhead and meta information, the efficiency for 100 data values does not improve much compared to 1 or 10 data values because the message must be split into multiple \gls{tcp} segments and thus multiple Ethernet packets. These issues are also responsible for the low performance in terms of update times.  However, what is an advantage for this interface is the direct access to the data without reconfiguration of the \gls{plc}.

\subsubsection{\gls{opcua} \gls{pubsub}}
\begin{figure}[htbp]
\centerline{\includegraphics[width=.7\columnwidth]{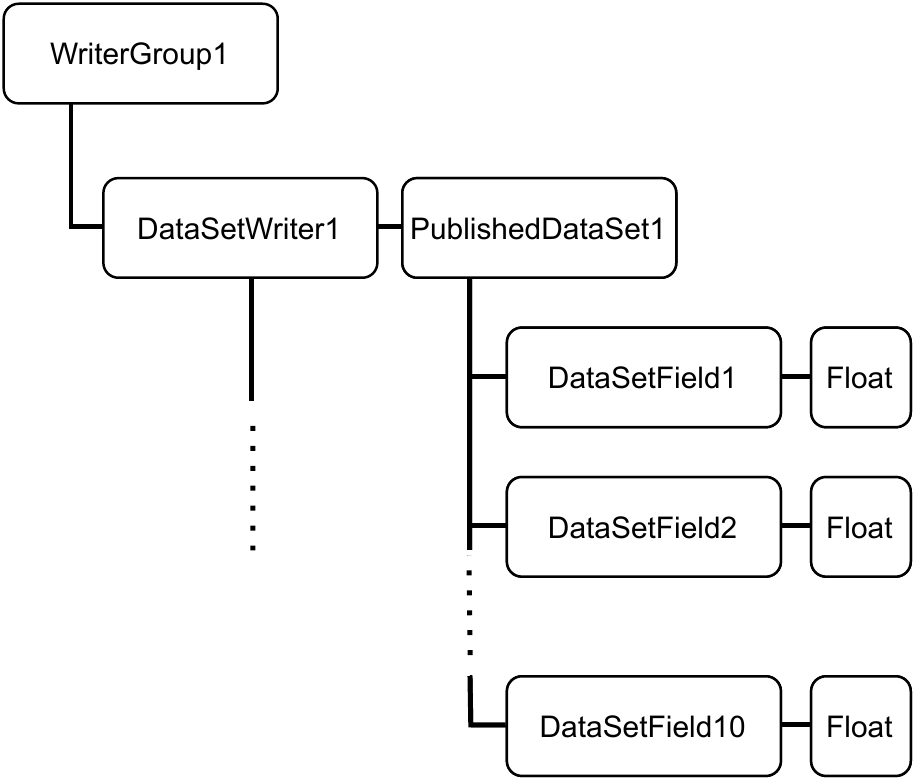}}
\caption{Structure of \gls{opcua} \gls{pubsub} network messages.}
\label{fig:pubsub}
\end{figure}

Due to the drawbacks of the server client model concerning protocol overhead and the descresed protocol efficiency of exchanging requests and responses, part 14 of the \gls{opcua} specifications adds the \gls{pubsub} pattern. This allows many subscribes to register for a specific content \cite{IEC625414}. For the message distribution both broker-based protocols, in particular \gls{mqtt} and \gls{amqp}, and UADP, a custom  \acrshort{udp}-based distribution based on the \acrshort{ip} standard for multicasting has been defined. Due to the advantages to send real-time messages on the field level directly on the data link layer, part 14 defines the transport of \gls{pubsub} messages based on Ethernet frames. Until now, the \glspl{plc} of the S7 family only support the data exchange with UADP. Therefore, only this protocol is discussed.
As shown in Fig. \ref{fig:pubsub}, there are several possibilites regarding structure of the packet. 
Since this interface requires reconfiguration of the \gls{plc}, it is not plug \& play capable. First a \textit{WriterGroup} must be defined. Under the \textit{WriterGroup} are logically grouped the so-called \textit{DataSetWriters}, which are associated with a \textit{PublishedDataSet}. Furthermore, the \textit{PublishedDataSet} contains the \textit{DataSetFields} where the specific data value is placed. Additionally, the \gls{opcua} data type of the data is added as meta information so that the subscriber can interpret the data correctly. We decided to configure only one \textit{DataSetWriter} and one \textit{PublishedDataSet} containing 1 to 10 data values for our measurements. So far, it is only possible to use 10 \textit{DataSetFields} per \textit{PublishedDataSet} and two \textit{DataSetWriters} per \textit{WriterGroup}. Therefore, the maximum data values that can be sent per \textit{WriterGroup} is 20. For this reason, we can only estimate the protocol efficiency and update time for 100 data values. Independently, \gls{opcua} \gls{pubsub} performs best in terms of update times for 1 and 10 data values of $\mathrm{\approx 1}$ms and 1.26 ms, respectively. Moreover, the achievable protocol efficiency is comparable to that of \gls{ouc}. This is the result of the low overhead given by the metadata. If this is even better for 1 and 10 data values, the additional overhead of 1 byte per data value for 100 data values compared to the \gls{tcp}-based \gls{ouc} is slightly higher than the \gls{tcp} protocol overhead including acknowledgements.

\subsection{Summary}

Looking deeper into the different interfaces, it can be seen that each of the interfaces has strengths in one of the categories. %
If a low overhead is the most important requirement, 
\gls{ouc} is best suited. If the focus is on plug \& play functionality to integrate brownfield devices without having to reconfigure the \gls{plc}, LIBNODAVE or the \textit{ReadService} of the \gls{opcua} server client model can be used. Here, a decision must be made between good performance and the need of meta information. 
Last but not least, \gls{udp}-based \gls{opcua} \gls{pubsub} provides a very good trade-off in terms of update time, protocol efficiency, and the presence of at least some meta information of the data values plus a standardized communication protocol. This is very important for the required interoperability of devices and applications on the way to Industry 4.0. Here also the combination of \gls{tsn} and the Ethernet-based \gls{opcua} \gls{pubsub} has already been discussed \cite{opcuart}. The use of raw Ethernet frames can save both the overhead and processing of the \gls{udp}/\gls{ip} protocol headers. Therefore, increased performance of this interface can be assumed. 
\section{Conclusion}%
\label{sec:Conclusion}
In this paper, we investigated open interfaces of two \glspl{plc} of Siemens S7 series. Therefore, we assessed all available interfaces using standard Ethernet according to qualitative aspects such as the protocol used, plug \& play capability, and the availability of metadata. Furthermore, quantitative criteria, such as protocol efficiency and update time, were evaluated for each of the interfaces of both \glspl{plc}. It turned out, that all of the interfaces, that were taken into account have their strengths and weaknesses. Concluding, the interface and protocol used must be carefully selected according to the requirements of the particular application.

\section{Future Work}%
\label{sec:Future Work}
Even though our research mainly focused on the best performance in terms of lowest overhead, least update time, most metadata, and plug \& play capability, other protocols may be attractive for various \gls{iiot} applications. Examples include application layer protocols such as \gls{mqtt}, \gls{coap}, and \gls{http}. Especially when dealing with 1-to-n relationships and latency or minimal update time are not the main concern, these protocols can be a good solution. 

In addition, it would be interesting to examine break-even points for a realistic computing offloading scenario, based on the chosen interface, network design, and complexity of the algorithm. For the latter, different levels of complexity are possible, such as classical \gls{pid} controllers, more complex ones, such as \glspl{lqr} in combination with Kalman filters or state observers for multiple states, or ones that solve nonlinear equations.




\balance
\printbibliography%

@INPROCEEDINGS{3gppts22104,
	AUTHOR			= {3GPP},
	TITLE			= {TS 22.104; 5G; Service requirements for cyber-physical control applications in
vertical domains},
	YEAR			= {2020},
   % URL				= {https://www.plattform-i40.de/I40/Redaktion/EN/Downloads/Publikation/structure-of-the-administration-shell.pdf?__blob=publicationFile&v=5},
   % NOTE			= {Last accessed: 2018-03-26},
}

@INPROCEEDINGS{IEC625411,
	AUTHOR			= {},
	TITLE			= {IEC 62541-1, OPC Unified Architecture - Part 1: Overview and concepts},
	YEAR			= {2010},
    month			= {},
    %URL			= {},
    %NOTE			= {Last accessed:},
}

@INPROCEEDINGS{IEC625414,
	AUTHOR			= {},
	TITLE			= {IEC 62541-14, OPC Unified Architecture - Part 14: PubSub},
	YEAR			= {2019},
    month			= {},
    %URL			= {},
    %NOTE			= {Last accessed:},
}

@INPROCEEDINGS{gundall20185g, 
author={M. Gundall and J. Schneider and H. D. Schotten and M. Aleksy and others},%D. Schulz and N. Franchi and N. Schwarzenberg and C. Markwart and R. Halfmann and P. Rost and D. Wübben and A. Neumann and M. Düngen and T. Neugebauer and R. Blunk and M. Kus and J. Grießbach}

@INPROCEEDINGS{ldd,
	AUTHOR			= {K. Pietrusewicz and L. Urbanski},
	TITLE			= {Control programming software strategies for industrial systems},
	YEAR			= {2011},
    month			= {},
    %URL			= {},
    %NOTE			= {Last accessed:},
}

@INPROCEEDINGS{8254628, 
author={S. Tayade and P. Rost and A. Maeder and H. D. Schotten}, 
booktitle={GLOBECOM 2017 - 2017 IEEE Global Communications Conference}, 
title={Device-Centric Energy Optimization for Edge Cloud Offloading}, 
year={2017}, 
volume={}, 
number={}, 
pages={1-7}, 
doi={10.1109/GLOCOM.2017.8254628}, 
ISSN={}, 
month={Dec},}

@inproceedings{opcuart,
author = {Pfrommer, Julius and Ebner, Andreas and Ravikumar, Siddharth and Karunakaran, Bhagath},
year = {2018},
month = {07},
pages = {},
title = {Open Source OPC UA PubSub over TSN for Realtime Industrial Communication},
doi = {10.1109/ETFA.2018.8502479}
}

@INPROCEEDINGS{gundall2020application,
  author={M. Gundall and D. Reti and H.~D. Schotten},
  booktitle={2020 IEEE 18th International Conference on Industrial Informatics (INDIN)}, 
  title={Application of Virtualization Technologies in Novel Industrial Automation: Catalyst or Show-Stopper?}, 
  year={2020},
  volume={1},
  pages={283--290},
  organization={IEEE}
  }

@INPROCEEDINGS{gundall2020introduction,
  author={M. Gundall and C. Glas and H.~D. Schotten},
  booktitle={2020 25th IEEE International Conference on Emerging Technologies and Factory Automation (ETFA)}, 
  title={Introduction of an Architecture for Flexible Future Process Control Systems as Enabler for Industry 4.0}, 
  year={2020},
  organization={IEEE},
    pages={1047--1050},
  volume={1},
  }

@ARTICLE{gundall2021introduction,
  author={M. {Gundall} and M. {Strufe} and H. D. {Schotten} and P. {Rost} and C. {Markwart} and R. {Blunk} and A. {Neumann} and J. {Grießbach} and M. {Aleksy} and D. {Wübben}},
  journal={IEEE Access}, 
  title={Introduction of a 5G-Enabled Architecture for the Realization of Industry 4.0 Use Cases}, 
  year={2021},
  volume={9},
  number={},
  pages={25508-25521},
  doi={10.1109/ACCESS.2021.3057675}}

@INPROCEEDINGS{gundall2021feasability,
  author={M. {Gundall} and C. {Glas} and  H. D. {Schotten}},
  booktitle={2021 IEEE International Conference on Industrial Technology (ICIT)}, 
  title={Feasibility Study on Virtual Process Controllers as Basis for Future Industrial Automation Systems}, 
  year={2021},
  volume={},
  number={},
 % pages={},
 % doi={}
 }

@ARTICLE{5535166,
  author={T. {Sauter}},
  journal={IEEE Transactions on Industrial Electronics}, 
  title={The Three Generations of Field-Level Networks—Evolution and Compatibility Issues}, 
  year={2010},
  volume={57},
  number={11},
  pages={3585-3595},
  doi={10.1109/TIE.2010.2062473}}

@article{leitner2006opc,
  title={OPC UA--service-oriented architecture for industrial applications},
  author={Leitner, Stefan-Helmut and Mahnke, Wolfgang},
  journal={ABB Corporate Research Center},
  volume={48},
  pages={61--66},
  year={2006}
}

@misc{hergenhahn2011libnodave,
  title={LIBNODAVE, exchange data with Siemens PLCs},
  author={Hergenhahn, Thomas},
  year={2011}
}

@article{langner2013kill,
  title={To kill a centrifuge: A technical analysis of what stuxnet’s creators tried to achieve},
  author={Langner, Ralph},
  journal={The Langner Group},
  year={2013}
}

@ARTICLE{6009198,
  author={T. {Sauter} and M. {Lobashov}},
  journal={IEEE Transactions on Industrial Informatics}, 
  title={How to Access Factory Floor Information Using Internet Technologies and Gateways}, 
  year={2011},
  volume={7},
  number={4},
  pages={699-712},
  doi={10.1109/TII.2011.2166788}}
\nl%

%
%
\end{document}